\documentclass[namedreferences]{SolarPhysics}
\usepackage[optionalrh]{spr-sola-addons} 
\usepackage{graphicx}        
\usepackage{color}           
\usepackage{url}             

\newcounter{steps}
\newenvironment{Steps}%
   {\begin{list}{\textsc{Step} \arabic{steps}. }{\usecounter{steps}%
           \setlength{\labelsep}{0pt}\setlength{\leftmargin}{0pt}%
           \setlength{\labelwidth}{0pt}%
           \setlength{\listparindent}{0pt}}}%
   {\end{list}}

\begin{document}

\begin{article}

\begin{opening}

\title{Constraining coronal heating: employing Bayesian analysis techniques to improve the determination of solar atmospheric plasma parameters}

\author{Sotiris~\surname{Adamakis}$^{1}$\sep
        Robert W.~\surname{Walsh}$^{1}$\sep      
        Anthony J.~\surname{Morton-Jones}$^{1}$
       }
\runningauthor{Adamakis et al.}
\runningtitle{Bayesian Analysis}

   \institute{$^{1}$ University of Central Lancashire
                     \url{sadamakis@uclan.ac.uk}, \url{ajmorton-jones@uclan.ac.uk}, \url{rwwalsh@uclan.ac.uk}
             }

\begin{abstract}
One way of revealing the nature of the coronal heating mechanism is by comparing simple theoretical one dimensional hydrostatic loop models with observations at the temperature and/or density structure along these features. The most well-known method for dealing with comparisons like that is the $\chi^2$ approach. In this paper we consider the restrictions imposed by this approach and present an alternative way for making model comparisons using Bayesian statistics. In order to quantify our beliefs we use Bayes factors and information criteria such as AIC and BIC. Two datasets \cite{ugarte2005, priest2000} are re-analyzed using the method described above. For the \inlinecite{ugarte2005} dataset, we conclude apex dominant heating as the likely heating candidate, whereas the \inlinecite{priest2000} dataset implies basal heating. Note that these new results are different from those obtained using the chi-squared statistic. For this we suggest that proper usage of Classical and Bayesian statistics should be applied in order to make safe assumptions about the nature of the coronal heating mechanisms. 
\end{abstract}

\end{opening}

\section{Introduction}
\label{intro}
Magnetically confined plasma loops are the fundamental building blocks of the solar atmosphere. Whole loop structures are observed over an extensive spectral range while extending over a large range of length-scales and dynamical time-scales. In particular, the Solar and Heliospheric Observatory (SOHO), the Transition Region and Coronal Explorer (TRACE) and now Hinode record loop-like features from small-scale brightenings lasting for tens of seconds to large-scale (of the order of the solar radius), apparently static loops that last for many hours.

Recent interest in loops has centred on the definitive determination via remote sensing of the basic parameter values within these features. One dimensional hydrostatic simulations of loop plasma ({\it e.g.} \opencite{per00}) produce a temperature ($T$) and density ($\rho$) structure along the loop that results from a balance between thermal conduction along the field-lines, optically thin radiative loss and a predefined coronal heating term. Generally, this profile extends from a cooler temperature at the loop base (or footpoint) area up to a hotter temperature at the loop apex. However, it has been demonstrated that the temperature gradient ($\frac{dT}{ds}$ where $s$ is the distance along the loop) at each point along this profile is very dependent on where the energy deposition preferentially occurs ({\it e.g.} \opencite{priest2000}). If the heat input is predominantly located at the base, the temperature in the ``coronal'' part of the loop will be relatively constant ({\it i.e.} $\frac{dT}{ds} \approx 0$). In comparison, if the energy is released near the loop apex, a significant temperature gradient ($\frac{dT}{ds} > 0$) travelling away from the apex can result. By observing the local temperature (and/or density) at successive spatial locations along a well-defined loop and then comparing the resulting profile with one generated from a one-dimensional hydrostatic model, then this could provide a means of constraining the possible preferred spatial location of the heating within that loop structure.

Confirming the dominant energy deposition position has been more difficult than first imagined, including for example the dataset introduced by \inlinecite{priest2000} (PDS from now on) which has been interpreted by separate authors as uniform \cite{priest2000}, base \cite{asch01} and apex \cite{reale02,mackay2000} heating. Other datasets examine the variation in density rather than temperature (\opencite{ugarte2005}; UUDS from now on). It is of vital importance to include, {\it e.g.}, emission measure observations and/or density values together with the temperature values. Nonetheless, only temperature or density values were available to us. Here we have reanalyzed specific data to examine whether or not more robust statistical techniques yield different results.

In many of the above investigations, the important step of model comparison between the hydrostatic simulation and the observed plasma parameter values is undertaken by employing a weighted chi-squared analysis. Possibly this is an adequate, ``quick-look'' approach to tackling the model comparison. However two statistical ``obstacles'' present themselves here: 
\begin{enumerate}
\item The precision of the temperature observations may not be sufficient to discriminate one heating function from another. As we shall see, substantive changes in the nature of the heating function can result in only subtle changes in the temperature profile of the loop. 
\item The current approach (as in PDS and UUDS) for comparing one heating function model with another is to minimize the well-known statistic: 
\begin{equation}
\chi^2 = \sum_{i=1}^n \frac {\left(T_i - \widehat{T}_i\right)^2} {\sigma_i^2}, 
\label{chi}
\end{equation}
where $n$ is the number of grid points, $T_i$ is the observed temperature at distance $s_i$ along the strand, $\widehat{T}_i$ is the theoretical prediction of the model concerning the temperature and $\sigma_i$ is the standard deviation of the observed temperature. The procedure, applied to the solar coronal loop heating problem, would be to isolate the correct heating functional form ({\it e.g.} base or apex heating with an exponential profile) based on its capability to furnish the minimum statistic value. 
\end{enumerate}

There are a number of deficiencies to this procedure. First of all, for any heating functional form, there is a continuous range of parameter values, with an infinite set of possible parameter combinations, each of which is a candidate model. Therefore, it is not possible to properly compare one family's performance in fitting the data with another ({\it e.g.} apex heating with basal heating using an exponential form), without resorting to selecting specific values of the parameters. Such an approach was employed by UUDS, where a grid method of equally spaced parameter combinations was used. Of course, the concern is that somewhere we have missed certain parameter values between grid points which could reverse the conclusion reached in the model comparison assessment. 

Furthermore, using a procedure such as the one described above, there is no straightforward way of telling if we have significant evidence of one heating function being superior to another. The issue here is that the minimum $\chi^2$ statistic approach is not well suited to model comparison problems, since it is primarily a goodness-of-fit statistic. 

Another point of interest is that the minimum $\chi^2$ statistic approach is only strictly valid under the normal errors assumption. As with the UUDS observations, this is often clearly not the case --- the error bars may be asymmetric. This means a minimum $\chi^2$ based assessment may not be reliable. 

Finally, merely taking the model with the minimum $\chi^2$ statistic does not tell us anything about the quality of the model. One can always improve a model fit by adding increasingly more parameters until, at the point of nonidentifiability, the model fit equates to all observation values exactly (``joins the dots''), thus producing a $\chi^2$ statistic value of zero, which would be then the model of choice based on the minimum $\chi^2$ statistic criterion. However, it is clear that what we have done in that case is not come closer to the true picture: rather an artefactual model has been constructed which is unlikely to reflect the true picture. This whole problem is one of {\it overfitting}, whereby model fit is apparently improved by adding increasingly more parameters, and is not taken into account by simply using the $\chi^2$ statistic. 

The practical way round these issues is to resort to a simulation approach. In this paper we describe the use of a Bayesian Markov chain Monte Carlo (MCMC) analysis to solar coronal loop data, which embeds hydrodynamic modelling techniques (\opencite{wbh95}; see Section \ref{model}) within a basic Metropolis-Hastings algorithm \cite{mrrt53,has70}. 
Section \ref{realdata} investigates some real coronal loop datasets (PDS and UUDS) and subsequently presents quantitatively based conclusions concerning the nature of the heating of the loops examined. A discussion on how this work can be further progressed can be found in Section \ref{discussion}.

\section{Statistical Methods}

\subsection{General Approach}
In principle, the provision of observational temperature data with distance from the base of the loop will yield information concerning the nature of the heating function. Our approach employs a Bayesian analysis of the data using MCMC techniques, which are increasingly being employed upon astrophysical datasets \cite{adamakis07}. For example, \inlinecite{gl92} use Bayesian statistics for periodic signal detections in a dataset relating to the arrival times of individual X-ray photons used for the unmasking of X-ray pulsars. \inlinecite{kd98} use MCMC methods for the reconstruction of differential emission measure distributions from a solar active region spectrum and the quiet Sun spectrum. \inlinecite{henriques08} explore the parameter space of a semi-analytic model built upon the millennium dark matter simulation and compare it with three different observational datasets, including the galaxy K-band luminosity function, B-V colours, and the black hole-bulge mass relation. \inlinecite{dunkley05} compare cosmic microwave background and large-scale structure data with both a pure adiabatic cosmological model and a mixed adiabatic/isocurvature cosmological model. 

The Bayesian approach incorporates prior information we may have on model parameters we are interested in with the observational data (likelihood) to form the updated or posterior information we have on the parameters. This uses Bayes' Theorem:
\begin{equation}
p (\mathbf{P}|\mathbf{T}) = \frac {p(\mathbf{T}|\mathbf{P}) p(\mathbf{P})} {p(\mathbf{T})}
\label{posterior}
\end{equation}
for observations $\mathbf{T}$ and parameter space $\mathbf{P}$, where $p(\mathbf{T}|\mathbf{P})$ is the likelihood function, $p(\mathbf{P})$ the prior distribution, $p(\mathbf{T})$ the marginal likelihood and $p (\mathbf{P}|\mathbf{T})$ the posterior distribution. The forms for the likelihood function applied in this paper are given in Section \ref{data_distribution}. The prior distribution may or may not reflect knowledge we may have on a parameter. If it does not we use a so-called noninformative prior. 

The posterior distribution can be simulated using MCMC techniques. In our case, we use the Metropolis-Hastings algorithm to draw parameter values for each parameter in our model from the posterior distribution. This is not a straightforward application of the Metropolis-Hastings algorithm however, as we need to convert proposed heating function parameter values into model temperatures, $\widehat{T}_i$, at each distance, $s_i$. Using single-variable updates, the established hydrodynamic modelling code is used, with input being the current parameter values (including the proposed value of the parameter under consideration at a given iteration), to produce the $\widehat{T}_i$. The $\widehat{T}_i$ can then be used to construct the likelihood of the temperature observations, $T_i$. A further complication arises because we must have physically sensible temperature profiles with distance for the loop. If the $\widehat{T}_i$ is not monotonically decreasing from the apex to the base, then we must reject this generated set of parameter values. Thus, with $k$ the number of parameters in the model, our approach can be summarised as: 
\begin{Steps}
\item For the $j$th parameter, with current parameter value $p_j$, generate the proposed value, $p_j^*$, from a proposal distribution.
\item Using the current set of parameter values, $(p_1,p_2,\ldots,p_j^*,\ldots,p_k)$, call the hydrodynamic code to generate the $\widehat{T}_i$.
\item Reject the proposal $p_j^*$ if the $\widehat{T}_i$ are not monotonically decreasing. If so retain the current value $p_j$ and go to \textsc{Step}~\ref{MH4}. Otherwise accept the new parameter value for the parameter $p_j$ with probability 
\begin{equation}
\alpha(p_j,p_j^*) = \frac{p((\mathbf{P}_{k-j},p_j^*)|\mathbf{T})q((\mathbf{P}_{k-j},p_j^*),(\mathbf{P}_{k-j},p_j))}{p((\mathbf{P}_{k-j},p_j)|\mathbf{T})q((\mathbf{P}_{k-j},p_j),(\mathbf{P}_{k-j},p_j^*))}, 
\end{equation}
where $\mathbf{P}_{k-j}=(p_1,p_2,\ldots,p_{j-1},p_{j+1},\ldots,p_k)$ and $q(\cdot,\cdot)$ the proposal distribution and go to \textsc{Step}~\ref{MH4}.
\item Move to the next parameter $p_{j+1}$ and repeat the process.  \label{MH4}
\end{Steps}
We follow the same procedure for multivariate Metropolis updates, with the only difference that instead of updating a parameter at each time, we update the whole set of parameters, {\it i.e} instead of $p_j^*$ we have $\mathbf{P}^*$. Although single-variable Metropolis updates seems to behave slightly better than multivariate Metropolis updates \cite{n03}, it can be very time consuming especially for a large parameter space. For this reason we have used multivariate Metropolis updates for the examples presented in Section \ref{realdata}. 

In this way, through thousands of iterations, the marginal posterior distribution for each parameter is built up. This distribution can then be used in many ways to assess the parameter values, {\it e.g.} by drawing up $95\%$ credible intervals. If this whole analysis is repeated for each heating function model, then model comparison techniques can be used to provide a quantitative assessment of the likelihood of one model over the other (see Section \ref{Model Comparison}). 


\subsection{Data Distribution}
\label{data_distribution}

The likelihood of the data should represent the way that our observations are distributed. This can change according to the way we gather the data; for example, this could be due to the instrument we use to gather the data or whether we have symmetric or asymetric error bars. This must be taken into account in the analysis. 

\subsubsection{Asymmetric and symmetric error bars}

In case the error bars are not symmetric we have to deal with non-symmetric distributions. One popular right skewed distribution with positive support is the Gamma distribution. In this case, the observations $T_i,~i=1, \ldots, n$, will have model likelihood function: 

\begin{eqnarray*}
p(\mathbf{T} | \mathbf{P}) & = & 
\prod_{i=1}^{n}{p(T_i|\mathbf{P})} \\
& = & \prod_{i=1}^{n} T_i^{\gamma_i - 1} \frac {\exp(-T_i / \delta_i)} {\Gamma(\gamma_i) \delta_i ^ {\gamma_i}} I(T_i \in \mathcal{S}_1),
\end{eqnarray*}
where $\mathbf{T} = (T_1, \ldots, T_n)$, $\gamma_i, \delta_i$ are the parameters of the Gamma distribution, $\mathcal{S}_1$ is the domain of $p(T_i|\mathbf{P})$,  $\mathbf{P}$ is the parameter vector and
\begin{equation}
I(T_i \in \mathcal{S}_1) = \left\{
\begin{array}{ll}
1, \quad $~if~$ T_i \geq 0 $~and~$ T_{i+1} \leq T_i $~from the apex to the base~$ \\
0, \quad $~otherwise~$
\end{array}
\right.
\label{indicator}
\end{equation}
is the indicator function of the temperature. 

On the other hand, if the data we collected give symmetric error bars, we should use a symmetric distribution. A Gaussian distribution will usually be most appropriate. Thus, the model likelihood function of $T_i$ will be:
\begin{eqnarray*}
p(\mathbf{T} | \mathbf{P}) & = & 
\prod_{i=1}^{n}{p(T_i|\mathbf{P})} \\
& = & \prod_{i=1}^{n} \frac{1}{\sqrt{2\pi}\sigma_i} \exp\left(-\frac{(T_i-\mu_i)^2}{2\sigma_i^2}\right) I(T_i \in \mathcal{S}_1), 
\end{eqnarray*}
where $\mathbf{T} = (T_1, \ldots, T_n)$, $\mu_i, \sigma_i$ are the parameters of the Normal distribution and the indicator function is the same as in Equation (\ref{indicator}).

\subsubsection{Interpretation of error bars}
It is important to define clearly the standard deviation of the data from the error bars using probabilistic arguments. For instance, if we gather temperature values and we believe with some probability $pr_i$ that these values lie in the range $(T_{L,i}, T_{U,i})$ then we have:
\begin{equation}
P(T_{L,i} \leq T_i \leq T_{U,i}) = pr_i, 
\label{prob}
\end{equation}
where ($T_{L,i} , T_{U,i})$ are the lower and upper points of the error bar for the $i$th grid point respectively. The variance of the data distribution is calculated by solving Equation~(\ref{prob}) with the acceptance that the observed $T_i$ are the mode of that distribution. In the case of the Normal distribution we have: 
\begin{displaymath}
\sigma_i=\frac {T_{U,i}-T_{L,i}} {2\Phi^{-1}\left[(1 + pr_i)/2\right]}, 
\end{displaymath}
where $\Phi(\cdot)$ is the Cumulative Distribution Function (CDF) of the Normal distribution with mean $0$ and standard deviation $1$. In the special case that $pr_i = 0.9973, i = 1, \ldots, n$ then $2\Phi^{-1}\left[(1 + pr_i)/2\right] = 6$ and we get the ``3$\sigma$'' belief. In the case of the Gamma distribution, Equation (\ref{prob}) is difficult to solve analytically, so we turn to numerical methods, {\it e.g.} Newton-Raphson \cite{gcsr03}. The mode of the data distribution is calculated by the temperature values that are proposed from the model. We can then calculate the parameters from knowledge of the mode and variance.


\subsection{Heating Function: Models and Parameters}
\label{model}

The one-dimensional plasma equations employed in this model are \cite{wbh95}: 
\begin{eqnarray}
\frac{\partial{\rho}}{\partial{t}} + \frac{\partial{(\rho\upsilon)}}{\partial{s}} & = & 0
\label{eq:motion}
\\
\frac{\partial{\upsilon}}{\partial{s}} & = & \frac{\partial}{\partial{s}} \left(T^{5/2}\frac{\partial{T}}{\partial{s}} \right) - b \left[Q(T) - H(s) \right]
\label{eq:model}
\\
\rho & = & \frac{1}{T}, 
\label{eq:density}
\end{eqnarray}
where $\rho$ is the density, $t$ is the time, $\upsilon$ is the velocity of the plasma, $s$ is the distance along the loop, $Q$ is the radiative loss function, $H$ is the heat input and $b = \frac{\rho_{c}^{2} \chi_c T_{c}^{\theta_c} l^2} {\kappa_0 T_{c}^{7/2}}$ with $\rho_c$ the critical value of the density, $l$ the total length of the loop, $\kappa_0=10^{-11}$ for the corona and $T_c$ the critical value of the temperature. We also assume gravity and viscosity negligible, $s, t, \upsilon, T, \rho$ are normalized and the pressure is assumed to be constant along the loop ({\it e.g.} the conductive velocity is much smaller than the sound speed). For the radiative loss we have employed a piecewise function of the form $Q(T) = \chi T^\theta$. For $\chi$ and $\theta$ we use the values that are provided in \inlinecite{hild74}; this is a standard approach in all loop modeling. Other forms of the radiation function could be introduced. It is important to realize that if these forms differ greatly from \inlinecite{hild74}, then this could have a profound effect on the resulting thermal profile and thus on the resulting statistically significant heating solution (the values of beta). 

However, over the apex temperature ranges considered here (about 2MK), the radiative loss function will have a value that is a few orders of magnitude less than the corresponding thermal conduction. One could envisage the theoretical situation where a loss function is introduced that has a much greater value at higher temperatures, comparable to the conductive losses. The heating function would have to ``compensate'' for this, producing a higher value at the apex, thus pushing a Hildner-derived basal heating solution towards a more spatially uniform case say. However, given the above relative estimates, we would contend that the form of the radiative loss function in this case has a small effect. The reader is referred to \inlinecite{wbh95} where effect of changing the radiative loss function for a time-dependent heat input was examined. 

It should be noted that there are limitations to using a one-dimensional model of the plasma equations along a loop. Here we assume a constant cross-sectional area which might not be the case (though the reader is referred to the loop cross-sectional work by \opencite{watko00}). Also, the observed loop might not be truly semi-circular, hence effecting the spatial temperature profile for model comparison. However, as a starting point for demonstrating the statistical analysis techniques to be employed, a one dimensional model is used noting that the above deficiencies should be examined further at a later stage.

Equations (\ref{eq:motion}) to (\ref{eq:density}) are solved with the following boundary conditions:
\begin{equation}
\upsilon=0, \frac {\partial{T}} {\partial{s}} = 0 \quad \mbox{at} \quad s = 0
\end{equation}
\begin{equation}
T = T_{foot} \quad \mbox{at} \quad s = 0.5, 
\end{equation}
where $s=0$ is the apex of the loop. 

Whereas the optically thin radiation loss function can be estimated from observations, the form of the heating function still remains a mystery. In the analysis that follows, we assume that the heating function $H(s)$ has the following general form:
\begin{equation}
\label{heat}
H(s) = \lambda \exp(\beta s). 
\end{equation}
Of course this is only one case. We can test different functions in order to see which one best describes our data (see implications of this in Section \ref{discussion}). Therefore, we start with an initial temperature that is constant along the loop and impose the heating function under consideration. The code is allowed to evolve until static equilibrium conditions are reached, {\it e.g.} $\upsilon \approx 0$. Then we can compare the temperature profile produced by the theoretical model with the observations. 

Applying Equation (\ref{heat}) to (\ref{eq:model}) we get:
\begin{equation}
\label{eq:fmodel}
\frac{\partial{\upsilon}}{\partial{s}} = \frac{\partial}{\partial{s}}\left(T^{5/2}\frac{\partial{T}}{\partial{s}}\right) - bQ(T) + \alpha \exp(\beta s), 
\end{equation}
where $\alpha = b \lambda$. We have replaced $b \lambda$ with $\alpha$ to improve the efficiency of the MCMC sampler, because $b$ and $\lambda$ will be extremely high posterior correlated. 


Thus, we have a range of parameters to investigate. Firstly, there is $\alpha$ and $\beta$ from the heating function. $\beta$ is exceptionally important because altering its value can have a profound effect on the nature of the heating profile. For example, if it is positive then more heat is deposited in the lower part of the loop (footpoint or basal heating). On the other hand, if $\beta$ is negtive then more heat is deposited in the upper part of the loop (apex heating). Note that when $\beta=0$ we have the ``unique'' case of uniform heating (although see discussion on this in Section \ref{discussion}). 

Secondly, we introduce $T_{foot}$ as an extra parameter. \inlinecite{ugarte2005} highlight the sensitivity of choosing this boundary condition. Finally, in our simplified HD equations, we assume an isobaric scenario. Thus, since pressure $p$ remains unchanged along the loop for a given set of parameter values, we have decided to leave its value floating. The pressure will always be equal to $p_c$. Given $p_c = \frac {R} {\tilde{\mu}} \rho_c T_c$, then if we assume we fix $T_c$ ($10^6$ K), then $\rho_c$ will be a changing value to explore. Subsequently, varying $\rho_c$ changes $b$ which hence becomes our fourth and final parameter. Please note that we assume that the length $l$ of the loop to be well known and thus simply defined. 

To sum up, let the parameter space be $\mathbf{P} = (b,\alpha , \beta, T_{foot})^T$ with observed temperature values $\mathbf{T} \in \mathcal{S}_1$ (the data). The values of $\mathbf{P}$ lie in the region $\mathcal{S}_2~\in~(0,\infty) \times (0,\infty) \times \mathcal{R} \times [0,\infty)$. The restriction $\alpha > 0$ is because extra heat should be added to the system, not subtracted from it. Methods for choosing priors are discussed in Section \ref{priors}.

\subsection{Model Comparison}
\label{Model Comparison}
One of the most well-known statistical methods for comparing between two statistical models is the likelihood ratio test (LRT). This test tends to give more weight to the more complicated model. As a result, it has been proposed to penalize the log-likelihood by a factor that includes the parameters of the model \cite{ak74,bhadown77}.  

For example, suppose we have to compare two hypotheses: $H_1$ and $H_2$. It is shown that the ratio between the maximized likelihood under each hypothesis follows a $\chi^2$ distribution with degrees of freedom the difference between the number of parameters in each model (under certain regularity conditions --- see \opencite{protas02} for more discussion). However this suffers from four important aspects: 1) the testing models should be nested ({\it e.g.} the allowed parameter values of one model must be a subset of the other --- see \opencite{cla01}), 2) certain regularity conditions should be met, 3) the most simple hypothesis should not be at the boundaries of the parameter space of the most complicated hypothesis, and, 4) the dataset should be large in order the likelihood ratio to be approximated by the $\chi^2$ distribution. In our analysis the first three requirements are met, but the fourth depends on the dataset we have. For this reason, we would prefer to use a model selection method that has a broader range of applications. Thus, we will turn to Bayes factors. 

A point in favour of Bayesian statistics is that we do not have to deal with the misconceptions of the $p-\mbox{values}$ with Classical statistics. For example, in Bayesian statistics data not observed and extreme values are irrelevant. Furthermore, the LRT like any test in Classical statistics can be used only to reject the null hypothesis. It can not provide evidence in favour of the null hypothesis. This is a usual misconception with Classical statistics because a failure to reject a null hypothesis does not provide evidence of its favour. 

In Bayesian statistics, Bayes factor is considered to be the traditional way of testing two or more hypotheses. Suppose $H_1, H_2$ are the two hypotheses we want to test. The odds form of Bayes's theorem is:
\begin{equation}
\frac {p (H_1|\mathbf{T})} {p (H_2|\mathbf{T})} = \frac {p (\mathbf{T}|H_1)} {p (\mathbf{T}|H_2)} \frac{p (H_1)} {p (H_2)} . 
\label{odds}
\end{equation}
Note that $p (H_k)$ is the belief we have about the truth of the hypothesis $H_k$ before we observe the data, $p (H_k|\mathbf{T})$ is what we get after we observe the data and $p (\mathbf{T}|H_k)$ is the marginal density, {\it i.e}. the belief of the data after we sum over the parameter space. The first term of the right hand side of Equation (\ref{odds}) is the \emph{Bayes factor}. The second term of the right hand side of Equation (\ref{odds}) is the prior odds of the two hypotheses. In the absence of any prior information for the two hypotheses we may assume this to be 1 ({\it i.e.} $p (H_1)=p (H_2)=0.5$), if the two hypotheses are exclusive. In that particular case Bayes factor is equal to the posterior odds. However, if we have some prior information about the hypotheses there is always the option to include it in the analysis. 

It is worth mentioning here that Bayes factor tends to be more sensitive to the choice of prior than the posterior probability of an interval \cite{kass93,kgr89} and so choice of priors becomes even more critical. This is because the marginal densities integrate over all possible values of unknown parameters in the models: 
\begin{equation}
p(\mathbf{T}|H_k) = \int_{\mathbf{P}} {p(\mathbf{T}|\mathbf{P}) p(\mathbf{P}) d\mathbf{P}}. 
\label{marginal}
\end{equation}
For example, if we use an improper prior (say Uniform with an infinite range) for a parameter of interest, this will result to ill-defined Bayes factors and posterior probabilities that prefer the simplest model, with probability one, regardless of the information from the data. This is widely known as the Bartlett's paradox \cite{bar57,lind57}. Apart from improper priors, this might be a consequence of using priors with a very large spread, in an effort to make our distribution non-informative, which in turn can lead to false conclusions \cite{kgr89}. Thus, in the case where there is no available prior information, the spread of the prior should not be very large in order to have effective results with Bayes factors. \inlinecite{raf96} propose a way to overcome this problem using the Laplace method for integrals. In this paper we present results from two procedures to estimate the marginal densities, $p (\mathbf{T}|H_k)$. Interested readers can follow up the descriptions of these methods in \inlinecite{kr95}, \inlinecite{rafb96} and \inlinecite{clyde07}, which describe these approaches in detail.

\subsubsection{Laplace estimator}
\label{laplace_estimator}

According to the Laplace estimator, we have the following approximation of a real-valued function $g(u)$ and $u$ a $l$-dimenssional vector:
\begin{equation}
\int {e^{g(u)}du} \approx (2 \pi)^{l/2} |A|^{1/2} \exp \left[ g(u^*) \right], 
\label{laplace}
\end{equation}
where $u^*$ is the value of $u$ that maximizes $g$ and $A$ is minus the inverse Hessian of $g(\cdot)$ evaluated at $u^*$. When Equation (\ref{laplace}) is applied to Equation (\ref{marginal}) we get the approximation: 
\begin{displaymath}
p(\mathbf{T}|H_k) \approx  (2 \pi)^{l/2} |\Psi|^{1/2} p(\mathbf{T}|\mathbf{P}^*) p(\mathbf{P}^*), 
\end{displaymath}
where $\mathbf{P}^*$ is the posterior mode of the parameters and $\Psi$ is minus the inverse Hessian of $\log \left [ p(\mathbf{T}|\mathbf{P}) p(\mathbf{P}) \right]$ evaluated at $\mathbf{P}^*$. Taking into account that $\Psi$ is asymptotically equal to the posterior covariance matrix, then the marginal likelihood is easy to be calculated after we have run the MH algorithm. Because of the fact that MCMC trajectories take occasional distant excursions, \inlinecite{lr97} suggest to use a robust estimator of the posterior covariance matrix like the weighted variance matrix estimate with weights based on the Minimum Volume Ellipsoid (MVE) estimate \cite{rvz90}. However, once we have removed the within-chain correlation, it would be valid to use the posterior covariance matrix. In Section \ref{realdata}, we will present estimation for both covariance matrices. Finally, we compare the values of the Bayes factors we get with the tables appeared in Section 3.2 of \inlinecite{kr95} for qualitative assessment.

\subsubsection{Importance sampling using Monte Carlo estimation}
\label{importance_sampling_estimator}

Alternatively, we can use an additional marginal density $q(\cdot)$ that places more mass in the ``important'' regions of the parameter space. Equation (\ref{marginal}) may be rewritten as:
\begin{displaymath}
p(\mathbf{T}|H_k) = \int_{\mathbf{P}} { \frac {p(\mathbf{T}|\mathbf{P}) p(\mathbf{P}) } {q(\mathbf{P})} q(\mathbf{P}) d\mathbf{P}}
\end{displaymath}
which will give for the estimate:
\begin{displaymath}
p(\mathbf{T}|H_k) = \frac {1} {m} \sum_{i=1}^{m} { \frac {p(\mathbf{T}|\mathbf{P}^{(i)}) p(\mathbf{P}^{(i)}) } {q(\mathbf{P}^{(i)})} }, 
\end{displaymath}
where $\mathbf{P}^{(i)}$ are sampled from the $q(\cdot)$ distribution. Of course the efficiency of this method will depend critically on the choice of the proposal distribution. 

In Section \ref{realdata} we have also tried the harmonic mean estimator \cite{geldey94}, but due to the fact that it suffers from infinite variance, it is far from convergence. 

$\\$

An alternative way of making model comparison is by using the Akaike Information Criterion (AIC) proposed by \inlinecite{ak74} and the Bayesian Information Criterion (BIC) proposed by \inlinecite{sc78}. The former proposes to choose the model that minimizes
\begin{displaymath}
\mbox{AIC} = -2(\textrm{log maximized likelihood}) + 2(\mbox{number of parameters}), 
\end{displaymath}
whereas the latter chooses the model that minimizes
\begin{displaymath}
\mbox{BIC} = -2(\textrm{log maximized likelihood}) + (\textrm{log}n)(\mbox{number of parameters}), 
\end{displaymath}
where $n$ in this case will be the number of the observed data points. AIC tends to be biased in favour of more complicated models, as the log-likelihood tends to increase faster than the number of parameters. BIC tends to favour simpler models than those chosen by AIC. Also BIC asymptotically provides the same results with Bayes factors and is considered as a crude approximation of Bayes factors. In the special case that we are dealing with a symmetric likelihood function (\emph{e.g.} Gaussian distribution) maximizing the likelihood function would be like minimizng the $\chi^2$ statistic. Thus, in the following we will also employ these tests.

\subsection{Choosing Priors}
\label{priors}

A natural way to address the prior distributions for our parameters is by considering their supports (in this case their domain). Thus, for $b, \alpha$ and $T_{foot}$ we would like a distribution which supports non-negative numbers and for $\beta$ a distribution which supports all the real numbers. Because of the fact that we do not have any prior information for the parameters $b, \alpha$ we will use improper priors for these two paramters ($p (b,\alpha) \propto 1$). Gamma and Normal distributions seem to be ideal for parameters $T_{foot}$ and $\beta$ respectively, {\it i.e.} $\beta \sim N(\beta_1,\beta_2^2)$ and $T_{foot} \sim Gamma(t_1,t_2)$, with the parameters $\beta_1,\beta_2,t_1,t_2$ yet to be defined. 

To avoid indeterminate Bayes factors, we decided to avoid using improper priors for the parameter of interest $\beta$. When we start the analysis it could be that we do not have any prior information at all. For example, the crucial parameter here is $\beta$. If there was not any prior information, we could use for estimation purposes a Normal prior distribution centered at $0$, {\it i.e.} $\beta_1=0$, with a big variation, {\it e.g.} $\beta_2^2=10^4$. This will make it ``non-informative'', but it will lead to Bartlett's paradox (see Section \ref{Model Comparison}) when we want to include model comparison into our analysis. Thus, it would be preferable to include any prior information available. For example, we set $99\%$ confidence on the parameter $\beta$ being between $-20$ and $20$. This means that we can assume $\beta_1=0, \beta_2=7.76$. Using similar notions we end up with $t_1=2, t_2=1$ (which will give $95.96\%$ for $T_{foot}$ to fall between $0$ and $5$ with mode at $1$). 

\begin{figure}    
   \centerline{\hspace*{0.015\textwidth}
               \includegraphics[width=0.515\textwidth,clip=]{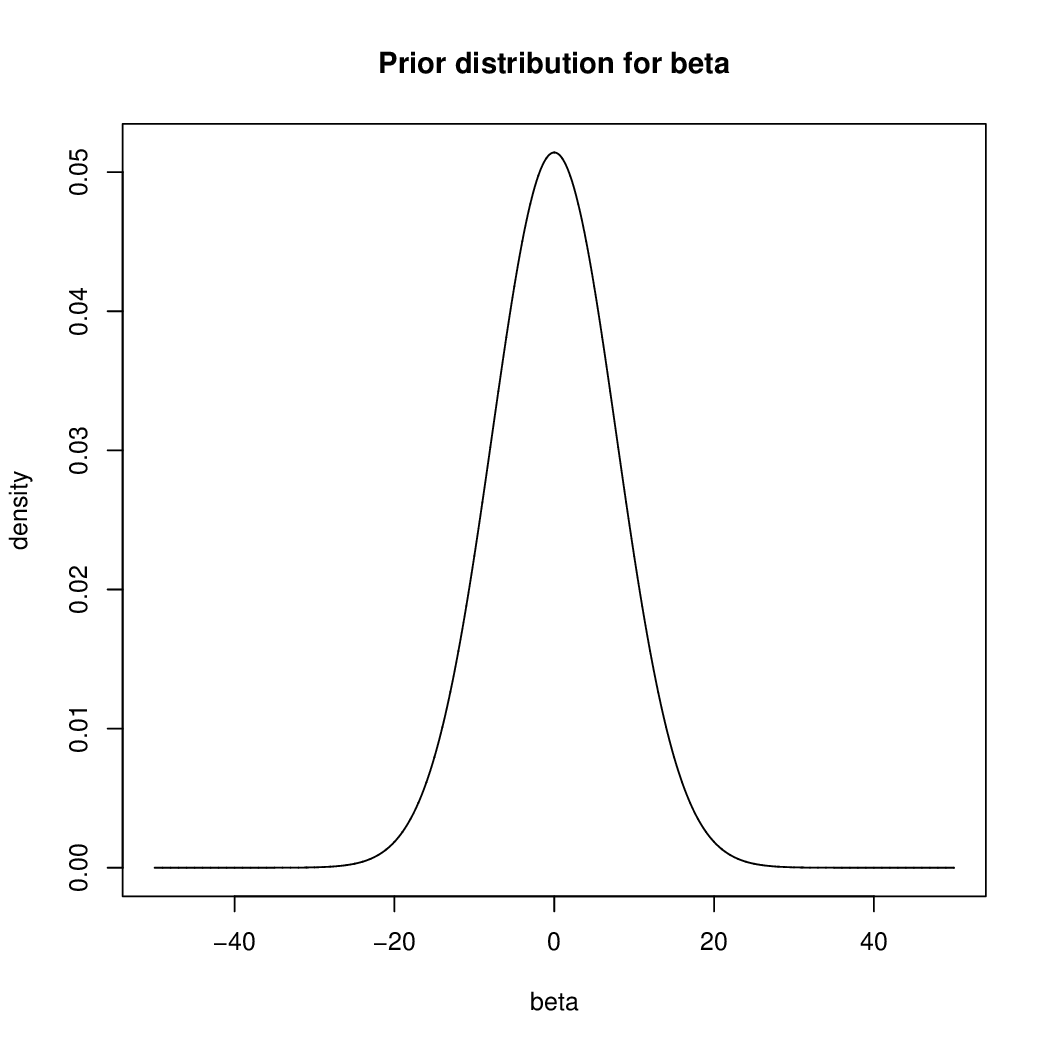}
               \hspace*{-0.03\textwidth}
               \includegraphics[width=0.515\textwidth,clip=]{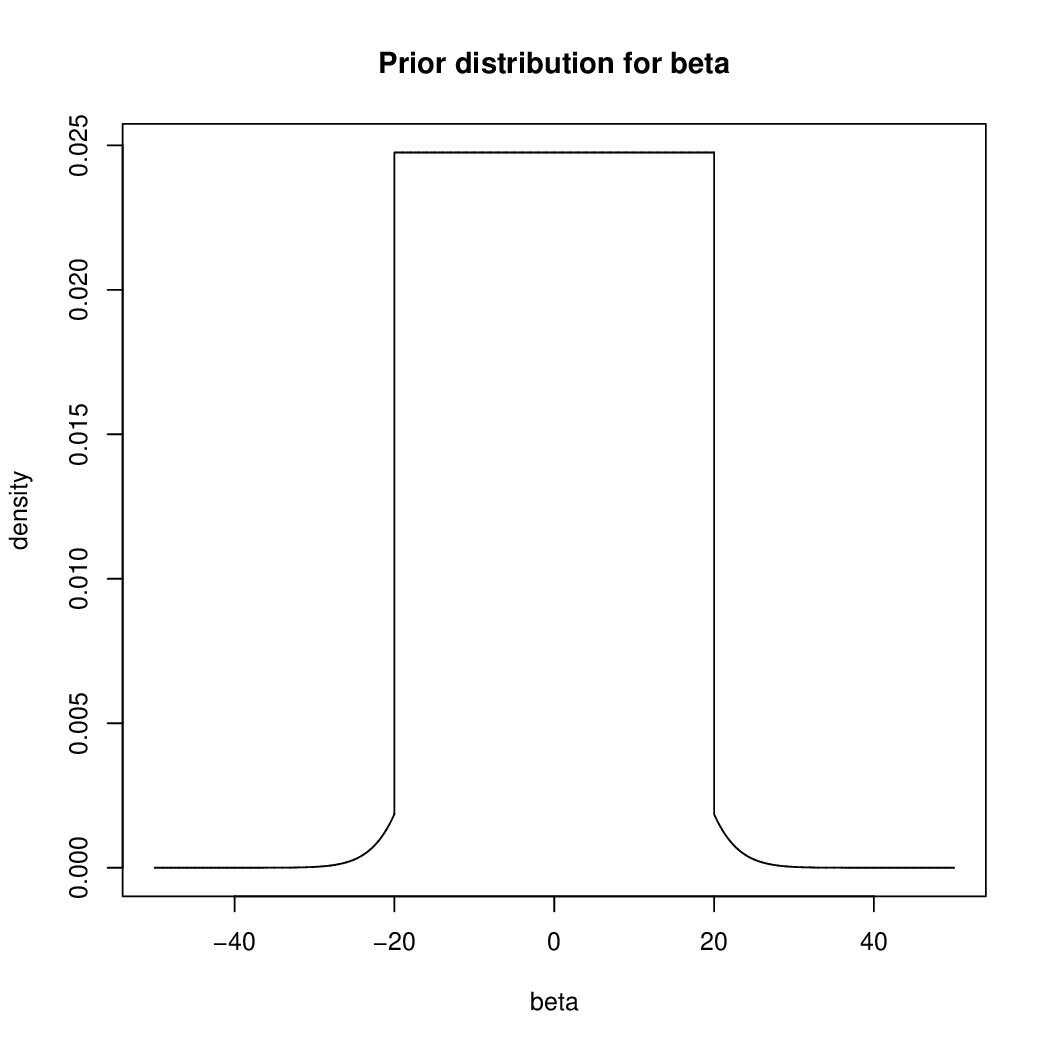}
              }
     \vspace{-0.35\textwidth}   
     \centerline{\Large \bf     
      \hspace{0.08 \textwidth}  \color{white}{(a)}
      \hspace{0.415\textwidth}  \color{white}{(b)}
         \hfill}
     \vspace{0.31\textwidth}    

\caption{(a) ``Common'' prior distribution for $\beta$. (b) ``Dual'' prior distribution for $\beta$.}
   \label{pdist}
\end{figure}

An alternative choice of prior distribution, based to the previous one, is by using ``Dual'' priors, which lead to well-defined Bayes factors, and can be described as following: choosing a Normal distribution with $\beta_1=0, \beta_2=7.76$ will give more weight into the values of $\beta$ that are closer to $\beta_1$. This means that $f(0)/f(20)=27.70$, where $f(\cdot)$ is the probability density function. So, instead of using a prior that will give more weight to certain values of the parameter, we may want to introduce a prior which will allow to jump between two values with equal prior probability. In that case the $p (\mathbf{P}) \propto 1$ described above seem ideal but has many problems. It seems natural to use a ``Dual'' prior that combines the above two distributions and integrates to unity. For example, for the $\beta$ parameter we may say that we are $99\%$ confident that the parameter $\beta$ should fall between $-20$ and $20$, as above, and we want the probability density function to be the same between $-20$ and $20$. This will give a probability density function as in Figure \ref{pdist}(b) instead of a probability density function as in Figure \ref{pdist}(a). ``Dual'' priors lead to well-defined Bayes factors. In Section \ref{realdata} we use ``Dual'' priors for parameters $\beta$ and $T_{foot}$. More specifically, for $\beta$ we use a uniform distribution between $-20$ and $20$ and a normal distribution with $\beta_1=0, \beta_2=7.76$ for every other value, whereas for $T_{foot}$ we use a uniform distribution between 0 and 3 and a Gamma distribution with $t_1=2, t_2=1$ for every other value. 

\subsection{Implementation of the MCMC Method}
\label{autocorr}
In generating our posterior distribution samples, the problem of within chain autocorrelation was found to be a significant problem. This problem means that much larger chains are required in order to achieve a representative sample from the target (posterior) distributions. There are several ways to deal with this problem, but we have found the most effective way has been to use the method described by \inlinecite{tm99}. The idea is to use more than one proposal in each step. This means that we start with a proposed parameter value combination. If they are accepted then move to the next step, otherwise propose another parameter value combination. If the second set is accepted then move to the next step, otherwise propose a third parameter value combination and so on. We can stop at any time we like this procedure, keep the current parameter values and move to the next step. The acceptance probability of each stage has to be adjusted in order to preserve a stationary distribution. This method has the advantage that we can test different proposals at each step, which can improve efficiency of mixing. We have used both two- and three-stages in our simulation procedures. At the first stage we propose values from an independent probability density. At the second stage (if needed) we propose values from a random walk probability density based on the current values. Finally, at the third stage (again if needed) we can propose values as in the second stage but with smaller standard deviation or from a random walk probability density based on the rejected values from the first stage. Furthermore, in order to improve the mixing of the chain, we reparameterize the space as our initial parameters are highly correlated. For the reparameterization and for the independent proposal of the first stage, we have run a pilot chain, \emph{i.e.}, we first run a simple Metropolis algorithm and from the crude estimates of the parameters we get we construct the independent proposal and choose the reparameterization scheme we will follow.

\section{Application of Observations}
\label{realdata}

In comparing the observed datasets with the HD simulation, we wish to examine the following four hypotheses: 
\begin{enumerate}
\item $H_1: \beta \neq 0$ --- heat input is not spatially uniform; 
\item $H_2: \beta = 0$ --- heat input is spatially uniform; 
\item $H_3: \beta > 0$ --- heat input is footpoint dominant; 
\item $H_4: \beta < 0$ --- heat input is apex dominant. 
\end{enumerate}
The method and the numerical codes were tested successfully against simulated observations. Due to limited space, these results are presented in \inlinecite{adamakis09}. In what follows, we present the results obtained from PDS and UUDS.

\subsection{Priest {\it et al.} Dataset}
\label{priestdataset}

\inlinecite{priest2000} analyze Yokhoh Soft X-ray observations from October 3, 1992 of a large coronal loop visible on the North-East limb of the large scale corona. Temperatures were obtained by a standard filter ratio technique. An image of this loop can be seen in Figure 9\emph{a} of \inlinecite{priest2000}. 

In this analysis we assume that the error bars reflect high degree of confidence. There are a number of important aspects to be kept in mind. Firstly, the loop under investigation is very long ($\approx 700$ Mm) yet the hydrostatic code we employ here ignores gravity, which really should be included. Secondly, there are important problems with how the observational results themselves are obtained. The structure widens as one travels from base to apex; therefore one cannot be sure that you are ``sitting'' on the same loop structure as one travels along the \citeauthor{priest2000} chosen data points. Thirdly, other papers question how the background emission has been extracted from the images. Thus, it could be regarded that this dataset is not very good example for this analysis. However, much interest has been generated by this paper and Bayesian analysis methods have never before been applied to this dataset. 

\begin{figure}    
   \centerline{\includegraphics[width=0.7\textwidth,clip=]{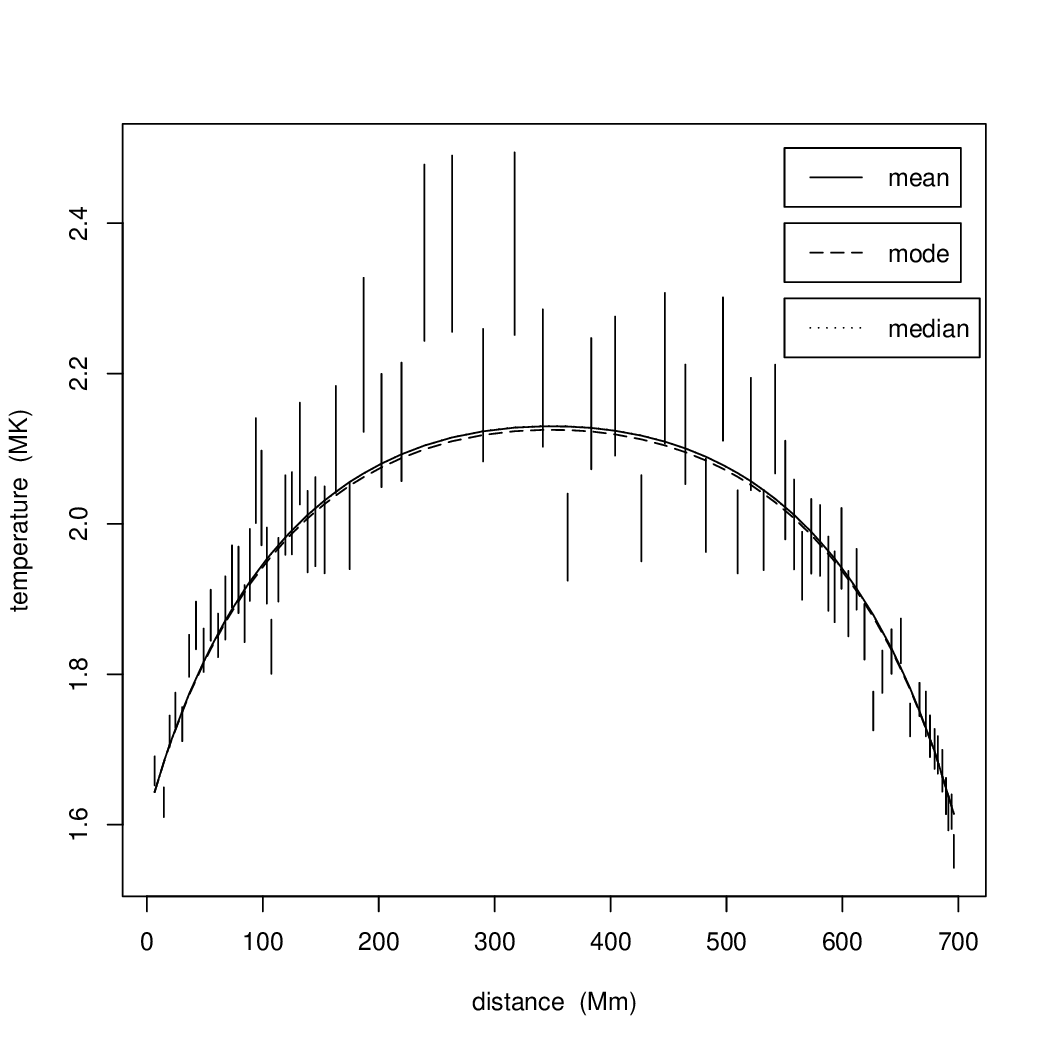}
              }
              \caption{Observational temperature values and fitted temperature profiles against distance along the loop for the PDS. The fitted temperature profiles are constructed using the mean (solid curve), joint mode (dashed curve) and median (dotted curve) values of the parameters taken from Table \ref{table3}.}
   \label{pr_model-data}
   \end{figure}

Since the only information of the data that we have are the error bars, we assume a Normal distribution for the data with $pr_i=0.98$, $i=1, \ldots, 74$. The summary statistics for the four parameters are presented in Table \ref{table3}. From Table \ref{table3} we can see that a $95\%$ credible interval for $\beta$ is between $1.58$ and $3.37$, which excludes negative values. In fact the probability of $\beta$ being negative is $P(\beta<0) \approx 0$. Thence, we would expect that $H_1$ and $H_3$ will give almost the same results. Since there is that high belief that $\beta>0$, it might seem pointless to construct Bayes factors. However, for the sake of completeness, we have calculated the values which will be useful for model comparison. Thus, according to Table \ref{table4} and using the Monte Carlo estimation with the probability density from stage $1$ of the delayed rejection algorithm of \inlinecite{mira01} as the additional probability density, we obtain the log-marginal density estimations $-137.59: -147.29: -137.52: -153.26$ for the hypotheses $H_1 : H_2 : H_3 : H_4$ respectively. According to \inlinecite{kr95}, this can be characterized as ``very strong'' evidence in favour of the $H_3$ hypothesis. AIC and BIC (see Table \ref{table44}) agree with the Bayes factors estimates and suggest that $H_1$ and $H_3$ are the best hypotheses. Therefore, we conclude that we have basal heating for this loop. This comes in contradiction with the \inlinecite{mackay2000} conclusion for this specific example. The three fitted curves of the mean, joint mode and median of hypotheses $H_1$ are depicted in Figure \ref{pr_model-data}.

\begin{table}
\caption{Summary of the posterior inference for $H_1$ for the PDS.}
\label{table3}
\begin{tabular}{ccccccc}     
  \hline                   
  & mean & mode & s.d. & $2.5\%$ & $50\%$ & $97.5\%$ \\
  \hline
$b$      & 41.38 & 3.11 & 32.68 & 1.32 & 33.34 & 117.73 \\
$\alpha$ & 15.74 & 12.17 & 3.33 & 10.77 & 15.21 & 23.43 \\
$\beta$  & 2.47 & 2.73 & 0.46 & 1.58 & 2.46 & 3.37 \\
$T_{foot}$ & 1.61 & 1.61 & 0.00 & 1.60 & 1.61 & 1.61 \\
  \hline
\end{tabular}
\end{table}

\begin{table}
\caption{Log-marginal likelihood estimates for the PDS. 1: Laplace method with posterior covariance matrix, 2: Laplace method with robust posterior covariance matrix, 3: Monte Carlo estimation with the probability density from stage $1$ as the additional probability density.}
\label{table4}
\begin{tabular}{ccccc}     
  \hline                   
  & $H_1: \beta \neq 0$ & $H_2: \beta =0$ & $H_3: \beta >0$ & $H_4: \beta <0$ \\
  \hline
1 & -136.35 & -146.49 & -136.39 & -151.70 \\
2 & -137.25 & -146.99 & -137.32 & -152.45 \\
3 & -137.59 & -147.29 & -137.52 & -153.26 \\
  \hline
\end{tabular}
\end{table}

\begin{table}
\caption{Information criteria for the PDS.}
\label{table44}
\begin{tabular}{ccccc}     
  \hline                   
  & $H_1: \beta \neq 0$ & $H_2: \beta =0$ & $H_3: \beta >0$ & $H_4: \beta <0$ \\
  \hline
AIC & 269.50 & 292.15 & 269.50 & 294.44 \\
BIC & 278.71 & 299.07 & 278.71 & 303.66 \\
  \hline
\end{tabular}
\end{table}

\subsection{Ugarte-Urra Dataset}
\label{uudataset}

\inlinecite{ugarte2005} examine SOHO CDS and EIT observations of a loop visible system on April 6, 1998 above the North-East limb at a latitude of $\sim 48^{\circ}$. The loop was visible in the hottest lines of the dataset, \emph{i.e.} Fe XVI 360.76 \AA\ ($\log(T) \sim 6.4$ K), Si XII 520.67 \AA\ and Fe XIV 334.17 \AA\ ($\log(T) \sim 6.3$ K). For cooler lines, \emph{i.e.} $\log(T) \le 6.2$ K, the apex is fainter and footpoints are more prominent. They also used spectral line ratio diagnostic techniques to determine the electron density along this off-limb coronal loop (see Figure 1 in \opencite{ugarte2005} for the specific CDS observation). 

Those authors used a 1D hydrostatic model similar to the one utilised in this paper to generate theoretical density profiles for comparison with the observations. Using a minimum chi-squared analysis, they concluded that the best fit, minimum chi-squared case resulted from a heating function that was weighted preferentially towards the loop base (see Figure 8 in \opencite{ugarte2005}). This density profile is reanalyzed here but now the model comparison step is undertaken using the Bayesian analysis method outlined in Section \ref{Model Comparison}.

\begin{figure}    
   \centerline{\includegraphics[width=0.7\textwidth,clip=]{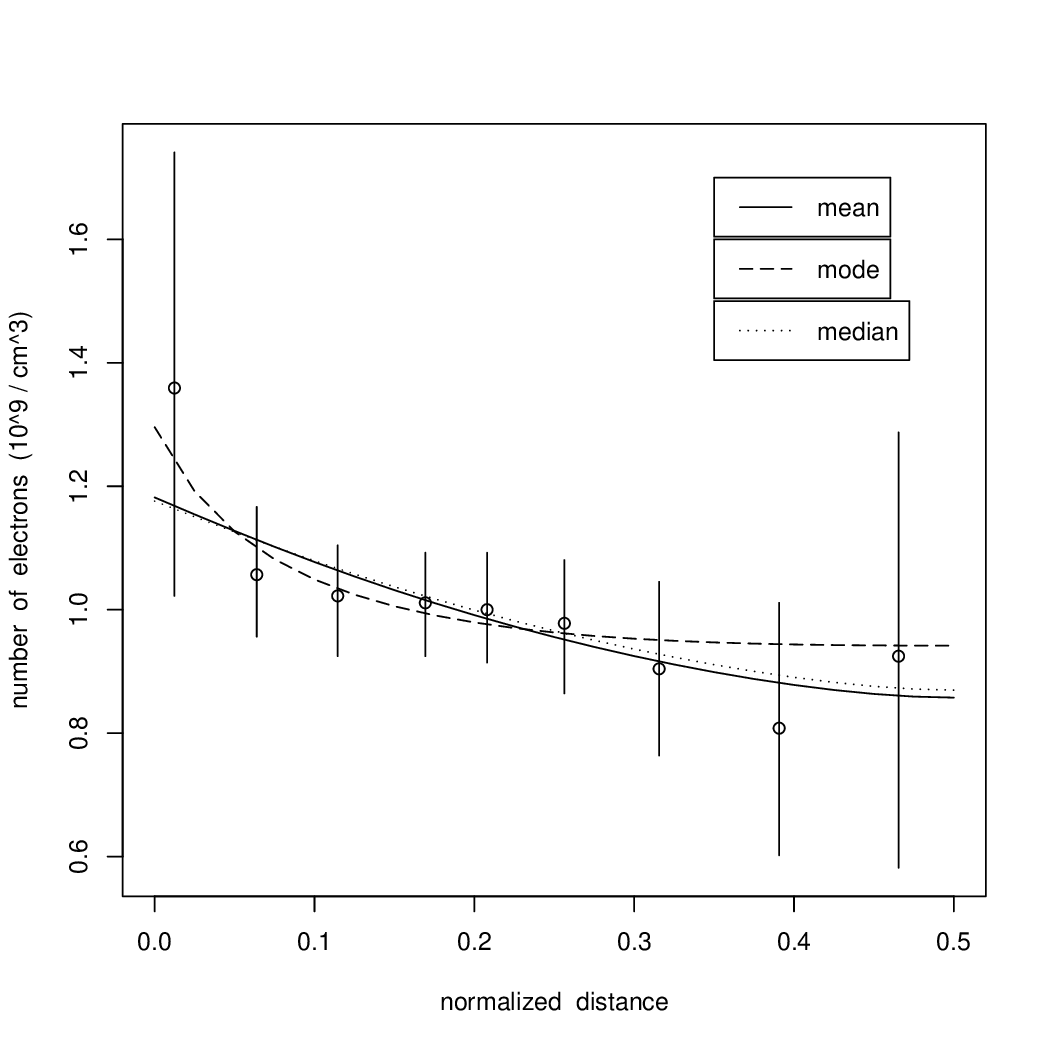}
              }
              \caption{Observational density values and fitted density profiles against distance along the loop for the UUDS. The fitted density profiles are constructed using the mean (solid curve), joint mode (dashed curve) and median (dotted curve) values of the parameters taken from Table \ref{table1}.}
   \label{rww_model-data}
   \end{figure}

Since the error bars are not symmetric we assume a Gamma distribution for the data with $pr_i=0.9973$, $i=1, \ldots, 9$. From the summary statistics in Table \ref{table1} we can conclude that the model prefers the negative values of $\beta$. To support this we have calculated the probability of $\beta$ being negative, {\it i.e.} $P(\beta < 0) \approx 0.90$. Table \ref{table2} shows the log-marginal likelihood estimates for each of the four hypotheses. All of the estimates suggest that the most probable hypothesis is $H_4$. For example, if we use the Laplace method with posterior covariance matrix, the log-marginal likelihood estimates are $13.94: 12.08: 11.32: 14.16$ for the hypotheses $H_1: H_2: H_3: H_4$ respectively. This can be characterized as ``positive'' evidence in favour of the $H_4$ hypothesis, according to \inlinecite{kr95}. The information criteria (see Table \ref{table22}) suggest the hypotheses (in preference order): $H_2, H_3, H_4$. Note that hypothesis $H_3$ is preferable than hypothesis $H_4$. This is because information criteria are focused only in the natural logarithm of the likelihood (plus the correction factor), ignoring the parameters dispersion under each hypothesis. Thus, we conclude to an apex heating mechanism. This result is at odds with the conclusion reached in \inlinecite{ugarte2005} as we shall discuss in the following section. If we consider only the hypothesis that maximizes the likelihood of the data, then hypothesis $H_3$ is the best, but when we add the correction factor (the factor that has to do with the number of parameters) hypothesis $H_2$ is much preferable! 
Figure \ref{rww_model-data} illustrates the three fitted curves of the mean, joint mode and median of hypothesis $H_1$ with the observed data points and error bars.

\begin{table}
\caption{Summary of the posterior inference for $H_1$ for the UUDS.}
\label{table1}
\begin{tabular}{ccccccc}     
  \hline                   
  & mean & mode & s.d. & $2.5\%$ & $50\%$ & $97.5\%$ \\
  \hline
$b$      & 9.02 & 4.42 & 7.50 & 0.31 & 7.11 & 27.52 \\
$\alpha$ & 33.61 & 2.47 & 19.16 & 4.15 & 32.40 & 76.42 \\
$\beta$  & -10.34 & 5.16 & 6.97 & -19.85 & -11.44 & 3.60 \\
$T_{foot}$ & 1.15 & 1.05 & 0.06 & 1.00 & 1.16 & 1.26 \\
  \hline
\end{tabular}
\end{table}

\begin{table}
\caption{Log-marginal likelihood estimates for the UUDS. 1: Laplace method with posterior covariance matrix, 2: Laplace method with robust posterior covariance matrix, 3: Monte Carlo estimation with the probability density from stage $1$ as the additional probability density.}
\label{table2}
\begin{tabular}{ccccc}     
  \hline                   
  & $H_1: \beta \neq 0$ & $H_2: \beta =0$ & $H_3: \beta >0$ & $H_4: \beta <0$ \\
  \hline
1 & 13.94 & 12.08 & 11.32 & 14.16 \\
2 & 12.85 & 11.87 & 10.25 & 13.26 \\
3 & 11.79 & 12.02 & 9.84 & 12.93 \\
  \hline
\end{tabular}
\end{table}

\begin{table}
\caption{Information criteria for the UUDS.}
\label{table22}
\begin{tabular}{ccccc}     
  \hline                   
  & $H_1: \beta \neq 0$ & $H_2: \beta =0$ & $H_3: \beta >0$ & $H_4: \beta <0$ \\
  \hline
AIC & -16.96 & -18.27 & -17.30 & -15.79 \\
BIC & -16.18 & -17.68 & -16.51 & -15.00 \\
  \hline
\end{tabular}
\end{table}


\section{Discussion and Future Work}
\label{discussion}

In this paper we have presented a new method for comparing observations with theoretical models for solar astrophysical datasets. Bayesian statistics are generally more powerful than a $\chi^2$ test for the reasons discussed in Section \ref{intro}. Simply choosing the model paradigm which happens to furnish the minimum $\chi^2$ value can be problematic. In particular, this Bayesian MCMC approach allows for a quantitative assessment of all the parameters simultaneously.  We have demonstrated how this can be applied in real datasets (Section \ref{realdata}). 

The results from our analysis of the PDS (Section \ref{priestdataset}) show conclusively that we have basal heating for that loop. The combination of 74 observations and narrower error bars compared with the UUDS mean that all diagnostic assessments strongly and unequivocally indicate this form of heat input. However, as it has been mentioned earlier in this paper, different authors have found different spatial forms of the heating function for this same loop. This is likely to occur because different analysis techniques could give different temperature profiles that do not resemble to each other. 

The results of our stochastic analysis of the UUDS (Section \ref{uudataset}) are contradicting in terms of the way we choose to analyze them. On the one hand, we have found that those techniques which use maximization to locate or estimate parameters (AIC or BIC) suggest a uniform heating mechanism ($\beta=0$). However, if we use an integral approach, {\it e.g.} Bayes factors, a negative value of $\beta$ is found. 

The resolution of this contradiction is probably found in the examination of the 95\% credible interval for $\beta$ which straddles 0. The marginal distribution of $\beta$ is negatively skewed, allowing the mode to positive, whilst the majority of the values are negative (the median value is negative). However, (\emph{i}) since the dataset consists only of nine datapoints, (\emph{ii}) that we trust the prior information we use, and (\emph{iii}) that the estimations of the marginal densities converge, we will choose inference from the Bayes factors. Thus, we suggest that apex heating is more dominant, according to the dataset we have. 

This leads onto another important point: we need more than one of these diagnostic assessment techniques described in this paper in order to be able to make a rounded reliable judgment on the nature of the heating mechanism. Our recommendation is to use the 95\% credible intervals, together with a maximization and integration based method. Use of say, the $\chi^2$ method alone or fitting by eye may lead to a false conclusion. 

The fact that information criteria seem to select $\beta=0$ (uniform heating) as the hypothesis of choice for UUDS may simply reflect a position of the error bars being too wide in comparison with the magnitude of $\beta$, which in turn determines the strength of basal or apex heating. Therefore, even if a loop is basally or apex heated, the size of $\beta$ may simply be too small to be ``detected'' by the available data, and, rather like the null hypothesis in classical statistical testing, a conclusion of ``uniform heating'' may be decided upon. Improved data may then come to a very different conclusion on the same loop!

It must be kept in mind that this model comparison is only as good as the analyzed data to which it is applied. In determining any spatial variation in the thermal/density structure along a loop and relating this to a model form, the main drivers are the number of observed data points along the structure under observation and the size of the error bar associated with each data point. Of course, it is the case that you would want to maximize one (the number of observations obtained) and minimize the other (to produce the smallest error bar). 



The reason for choosing the Gamma and Gaussian distributions for the likelihood functions in Section \ref{data_distribution} is that they are the most well-used. Of course, non-symmetric error bars could be analyzed with a different data distribution, \emph{e.g.} a Log-Normal distribution. However, the error distribution is an unfortunate assumption we have to make. In order to justify which error distribution to choose we need a lot more data (from other loops) and will need to undertake an analysis that is outside of the aims of this paper. The sensitivity in the choice of data distribution is very interesting and will need further research. 

Given that the spatial resolution of new (future) instrumentation are (will be) an improvement on that considered in this paper, it is likely that the number of data points along a structure will not be a vital issue, assuming one is dealing with a loop of reasonable lengh ($>100$Mm say). For example, the spatial resolution of Hinode/EIS is over twice that of SOHO/CDS. Similarly, a decrease in the size of the associated error bars should occur with greater instrument sensitivity --- however, it is possible that with greater spatial resolution, longer exposure times may be required. 

With this in mind, future work in this area will include examining the density structure along many loop examples observed by Hinode/EIS. Also, the numerical scheme will be extended to include gravity, relevant to longer loops.

\appendix 

\renewcommand{\theequation}{A\arabic{equation}}
\setcounter{equation}{0}         
\renewcommand{\thefigure}{A\arabic{figure}} 
\setcounter{figure}{0}           
\renewcommand{\thetable}{A\arabic{table}} 
\setcounter{table}{0}            

The main scope of this paper is to analyze already published data values using Bayesian methods. For this reason the same temperature values as in \inlinecite{priest2000} and \inlinecite{ugarte2005} were employed. Reanalyzing these datasets is beyond the scope of this paper but will be considered by the authors in future work. 

A simpler approach can be to use both Frequentist and Bayesian methods. This would work as follows:
\begin{enumerate}
\item First apply the common LRT to check which of $H_1$ or $H_2$ is preferable. As a reminder, LRT theory suggests that as the sample size $n \rightarrow \infty$, then \linebreak $-2 \log \Lambda \sim X^2_1$, where:
\begin{displaymath}
\Lambda = \frac{sup\{ L(\mathbf{P}|\mathbf{T}):\beta = 0\}}{{sup\{ L(\mathbf{P}|\mathbf{T}):\beta \neq 0\}}},
\end{displaymath}
with $L(\cdot|\cdot)$ the likelihood function and the difference between the dimensions of $H_1$ and $H_2$ to be one. 
\item If $H_2$ is preferable our analysis is complete and we have identified uniform heating as the most probable regime. If $H_1$ is preferable then we can integrate the posterior probability of $\beta$ to check which of $H_3$ or $H_4$ is more likely to be true. 
\end{enumerate}

\section{PDS}

For the PDS, $-2 \log \Lambda = 24.69$, which makes the $p-\mbox{value}$ of this test to be $6.73 \times 10^{-7}$. In this case the $H_2$ hypothesis is rejected, which means that the parameter $\beta$ is statistically significant. Then from the posterior distribution of $\beta$, $P(\beta>0) \approx 1$ is obtained, which again suggests that basal heating is more dominant. Note that this approach gives a similar result with the Bayes factor.

\section{UUDS}

For the UUDS, a value of 0.69 is obtained for $-2 \log \Lambda$, which gives a $p-\mbox{value}=0.41$ for this test. However, the approximation required for the $\chi^2$ test must be questioned as the number of datapoints is limited ({\it i.e.} only nine). In this case the evidence provided by the data is not enough to reject $H_2$. Note that this approach is in accordance with the information criteria.

\begin{acks}
SA has been supported by an STFC grant. We are pleased to thank the anonymous referees for comments that helped improve the manuscript. 
\end{acks}


\bibliographystyle{spr-mp-sola}

\bibliography{biblio}

\end{article}

\end{document}